\documentclass[aps,prb,twocolumn,showpacs]{revtex4}
\usepackage{amsmath}
\usepackage{graphicx}
\usepackage{dcolumn}% Align table columns on decimal point
\usepackage{bm}% bold math

\newcommand\scalemath[2]{\scalebox{#1}{\mbox{\ensuremath{\displaystyle #2}}}}

\begin{document}

%\frontmatter
\title{
Nonreciprocal Dispersion of Spin Waves in Ferromagnetic Thin Films Covered with a Finite-Conductivity Metal }
\author{M. Mruczkiewicz, M. Krawczyk}
\affiliation{Faculty of Physics, Adam Mickiewicz University in Poznan, Umultowska 85, Pozna\'{n}, 61-614 Poland}
%\author{author1}
%\affiliation{university}

\date{\today}
\begin{abstract}
We study the effect of one-side metallization of a uniform ferromagnetic thin film on its spin-wave dispersion relation in the Damon-Eshbach geometry. Due to the finite conductivity of the metallic cover layer on the ferromagnetic film the spin-wave dispersion relation may be nonreciprocal only in a limited wave-vector range. We provide an approximate analytical solution for the spin-wave frequency, discuss its validity and compare it with numerical results. The dispersion is analyzed systematically by varying the parameters of the ferromagnetic film, the metal cover layer and the value of the external magnetic field. The conclusions drawn from this analysis allow us to define a structure based on a 30 nm thick CoFeB film with an experimentally accessible nonreciprocal dispersion relation in a relatively wide wave-vector range.\end{abstract}

\maketitle

\section{Introduction}
The recent development of nanotechnology in combination with the increased interest in the properties of artificial crystals brought back the topic of spin-wave (SW) nonreciprocity and its potential application in functional devices. Spin-wave nonreciprocity in thin plates of magnonic crystals (MCs) has been obtained recently by special arrangement of magnetic dots in a two-dimensional array saturated along the normal to the array plane\cite{Verba13} or by covering a one-dimensional MC with a metallic layer.\cite{Saratov,Saratov2,mru} Structures with a metallic overlayer have already been investigated for applications. A magnonic crystal consisting of an yttrium iron garnet (YIG) film with periodic metal stripes atop has been shown to act as a sensitive magnetic field detector.\cite{Inoue} Magnetic metal structures have been studied also in terms of soliton excitation,\cite{Ustinov} pumping\cite{pumping} and current-driven tunability.\cite{Polushkin} Though not considered in these investigations, the nonreciprocal properties of SWs can be exploited for functionality enhancement or development. Also, logic devices based on nonreciprocal properties of SWs can be designed, and amplitude modulation used for logic operations.\cite{logic} Hence the need for the elucidation of the mechanisms behind the nonreciprocal properties of SWs and the description of the parameters on which they can depend. 

The analytical approach used for calculating the SW dispersion relation in uniform thin films in the magnetostatic linear approximation is based on the definition of the dynamic components of the magnetization vector and the magnetic field in terms of the magnetostatic potential from the Maxwell and Landau-Lifshitz (LL) equations.\cite{DE} The magnetostatic potential is defined separately  for each layer (i.e., for the ferromagnetic material and its nonmagnetic surrounding) and combined with the electromagnetic boundary conditions. This leads to a system of secular equations which can be solved analytically in some cases. The dispersion relation of SWs in a magnetic material surrounded by a dielectric, in contact with a perfect conductor\cite{Sachardi} or separated from a perfect conductor by a dielectric of finite width can be determined in this way.\cite{Kawasaki} In a similar manner the dispersion relation has been obtained for SWs propagating in various geometries\cite{DE, De_Wames_2} with the exchange interaction taken into account,\cite{Sparks, De_Wames_3} or in a ferromagnetic film surrounded by a magnetic wall.\cite{lokk} The influence of the finite conductivity of the magnetic film on the SWs was studied with the use of Green’s functions.\cite{Mills} The phenomenological propagation loss theory is described in Ref.~[\onlinecite{Stancil}].

The configuration in which the external magnetic field $H_{0}$ is applied in the plane of the thin film and the wave vector of the propagating SW is perpendicular to this field is referred to as the Damon-Eshbach (DE) geometry.\cite{DE} For a uniform ferromagnetic film covered with a layer of a finite-conductivity metal an analytical solution has been obtained for this geometry with the field $H_{0}$ regarded as a function of the wave vector,\cite{De_Wames, De Wames 4} since in this case the secular equation leads to a quadratic polynomial formula for $H_{0}$. Here, a real set of parameters leads to a complex value of $H_{0}$. The analytical solutions for the SW frequencies $f$ are more complicated; numerical solutions can be obtained, though.\cite{Berg, Okamura 1, Okamura 2} The finite-difference time-domain method has been used for studying the broadband ferromagnetic resonance response of single-layer and bi-layer magnetic films in contact with a metal.\cite{Kostylev 2013 2} Other propagation and field configurations\cite{Volume, Yukawa} and more complex structures, such as multilayer systems\cite{Mills2,Suk} and MCs in contact with a perfect electrical conductor, have been studied as well.\cite{PWM, mru}

The dispersion of SWs in thin-film structures can be measured by different experimental techniques, in transmission with the use of microwave antennas \cite{Saratov} (also in magnetic structures in contact with a metal \cite{Saratov2,Okamura 1, Okamura 2}) or, since recently, by Brillouin light scattering (BLS).\cite{BLS1,BLS2} The resolution of the BLS technique has proved adequate for the study of nanosize films, with a $k$ step of the order of $0.5 \times 10^6$ 1/m.\cite{adayeye, Car} Especially with reference to these experiments it is interesting to define a structure with nonreciprocal properties of SWs available to such measurements. We will accomplish this by calculating the real and imaginary parts Re$(f)$ and Im$(f)$ of the frequency as a function of the wave number $k$ and other parameters. In order to find $f$ we will derive the secular equation for structures in which a metal of finite thickness and conductivity is placed on top of a homogeneous ferromagnetic film and can be separated from the ferromagnetic film in a controlled way. For this purpose we implement the LL and Maxwell equations in the finite element method (FEM) and solve the obtained equations by iterative methods. The FEM has already been used successfully for calculating the dispersion of SWs in thin films.\cite{adayeye,mru, mru_2}

The paper is organized as follows. In Section \ref{Sec:Method} we present the FEM and a semi-analytical method for the solution of the LL and Maxwell equations in the magnetostatic approximation for a magnetic film covered with a metal layer. In Section \ref{Sec:approximate} we analyze the influence of the conductivity to arrive at an analytical formula for the real part of the frequency in a magnetic film in direct contact with a metal. In Section~\ref{Sec:SB} we provide a detailed analysis focused on maximizing the nonreciprocal effect of the metallization on the dispersion relation. The conclusions are presented in the closing Section~\ref{Sec:Concl}.

\section{Method}\label{Sec:Method}

\begin{figure}[!ht]
\includegraphics[width=0.45\textwidth]{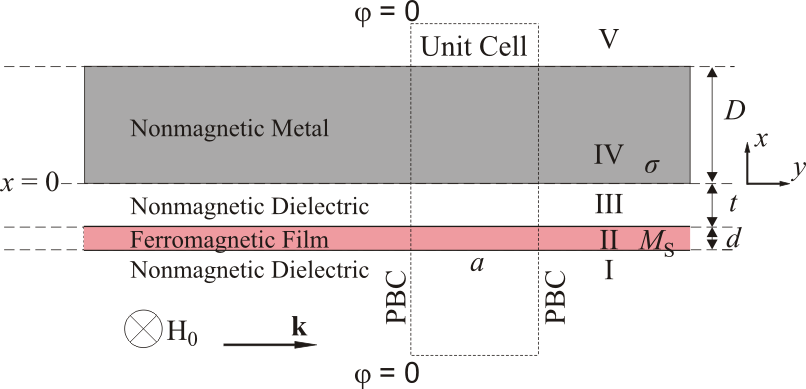}
\caption{(Color online) A structure under consideration. The bias magnetic field $H_0$ is in the film plane and directed along the $z$ axis. The SWs propagate along the $y$ axis. The rectangular unit cell used in numerical calculations is marked by dashed line. The PBC are used along the $x$ axis. The bottom and top border of the unit cell is far from the structure, at these borders $\varphi = 0$.}
\label{geometry}
\end{figure}

The geometry of the structure used in the calculations is shown in Fig.~\ref{geometry}. The structure consists of five regions. Region I and V are nonmagnetic dielectric surroundings, which are assumed to be infinite thick in semi-analytical calculations [described in Sec.~\ref{Sec:SA}] or to be very thick in finite element calculation  [see Sec.~\ref{Sec:FEM} for details]. Region II is a thin ferromagnetic film, characterized by the saturation magnetization $M_{\text{S}}$ and thickness $d$. The region IV is a nonmagnetic metal characterized by the conductivity $\sigma$ and thickness $D$. Region III is a dielectric nonmagnetic spacer, which separates the magnetic film from the metal overlayer and has thickness $t$. The external magnetic field $H_{0}$ is in the plane of the film--along $z$ axis and the wave propagation is assumed to be along $y$ axis, i.e., we consider only DE geometry. The origin of the Cartesian coordinate system is placed at the bottom surface of the metallic film.

 In calculations we will use the Maxwell equations in the magnetostatic approximation:
%\begin{eqnarray} \label{eq:1}
% \nabla \cdot ({\textbf h} + {\textbf m})= 0
%\end{eqnarray}
\begin{eqnarray}\label{eq:2}
 \nabla \times  \textbf{h}({\bf r}) &=& \sigma_{n} \textbf{e}({\bf r}), \\
 \nabla \times \textbf{e}({\bf r}) &=& -i \mu_{0} \omega ({\textbf h}({\bf r}) + {\textbf m}({\bf r})), \label{eq:3}
\end{eqnarray}
where $\mu_{0}$ is the permeability of vacuum, ${\bf r}$ is the position vector and $t$ is time. ${\textbf e}$ is the electric field vector, ${\textbf h}$ and ${\textbf m}$ are dynamic magnetic field and dynamic components of the magnetization vectors, respectively. The conductivity $\sigma_n$ is different from 0 only in the region IV, i.e., $\sigma_{n = \text{IV}} \equiv \sigma$. In the assumed geometry only $z$ component of the electric field is related to dynamic magnetic field, ${\bf e} = (0,0,e_{z})$. We will look for solutions in the form of monochromatic waves:
${\bf m}({\bf r},t) \sim \exp (i \omega t)$ and ${\bf h}({\bf r},t) \sim \exp (i \omega t)$, where $\omega = 2 \pi f$.
Maxwell equations are complemented with the LL equation of motion in the ferromagnetic film (in the region II):
\begin{eqnarray}
-i \omega  {\bf m}({\bf r}) = \gamma \mu_{0}
{\bf M}({\bf r},t) \times {\bf H}_{\text{eff}}({\bf r},t), \label{1}
\end{eqnarray}
where  $\gamma$ is a gyromagnetic ratio. ${\bf H}_{\text{eff}}$ denotes the effective magnetic field acting on the magnetization and is defined as ${\bf H}_{\text{eff}}({\bf r},t)=H_{0} \hat{z} +\nabla \cdot (\frac{2 A_{\text{ex}}}{\mu_{0}M_{\text{S}} }) \nabla {\bf m}({\bf r},t)+\textbf{h}({\bf r},t)$, where $A_{\text{ex}}$ is the exchange constant, $M_{\text{S}}$ is the saturation magnetization. We assume that the bias magnetic field $H_0$ is strong enough
to saturate the sample along the $z$ axis, thus  the magnetization vector ${\bf M}$ in linear approximation can be decomposed into the static and dynamic parts, parallel to the $z$-axis and laying in the plane ($x,y$), respectively: ${\bf M}({\bf r},t)= M_{0} \hat{z}+\textbf{m}({\bf r},t)$. We will assume $M_0 \equiv M_{\text{S}}$, $\textbf{m}=(m_{x},m_{y},0)$ and $\textbf{h}=(h_{x},h_{y},0)$.

\subsection{FEM}\label{Sec:FEM}
To find the dynamical components of the magnetization vector and the dispersion relation of SWs we will use FEM.  Artificial periodic boundary conditions (PBC) are imposed  at surfaces parallel to the $x$ axis at the borders of the unit cell of width $a$ as shown in Fig.~\ref{geometry}. Use of PBC allows us to define film of infinity length in numerical calculations. The solutions of the coupled LL and Maxwell equations in the periodic system can be written according to the Bloch theorem as: $\varphi(x,y)=\varphi^\prime(x,y) \text{e}^{i  k y }$, where $\varphi^\prime=m_{x}^\prime$, $m_{y}^\prime$, $h_{x}^\prime$, $h_{y}^\prime$ and $e_{z}^\prime$ are periodic functions of $x$ and $k$ is a wavenumber.

We expect that dynamic magnetic and electric fields should vanish at $x \rightarrow \infty$. To mimic this condition in FEM,  the Dirichlet boundary condition ($\varphi^\prime=0$) was used at the bottom and top border of the computational unite cell, which is far away (150 $\mu$m from the ferromagnetic film) from the magnetic film [Fig.~\ref{geometry}].  The boundary conditions for dynamic components of the magnetization vector are imposed only via LL and Maxwell equations, these provide  zero of the first derivative of the dynamical magnetization with respect to the normal to the surface.\cite{PK, mru_2,Sok}

After applying the Bloch theorem in linearized LL equation and in Maxwell equations the following set of  equations is formed:
\begin{widetext}
\begin{equation}
\left(
 \scalemath{0.75}{
 \begin{array}{ccccc}
\frac{i 2 \pi f}{\gamma \mu_{0}}& \nabla \cdot (\frac{2 A_{\text{ex}}}{M_{\text{S}} \mu_{0}}) \nabla +2 i k \partial_{y} -(\frac{2 A_{\text{ex}}}{M_{\text{S}} \mu_{0}}) k^2-H_{0}  & 0 & M_{\text{S}}& 0 \\
 - \nabla \cdot (\frac{2 A_{\text{ex}}}{M_{\text{S}} \mu_{0}}) \nabla - 2 i k \partial_{y} + (\frac{2 A_{\text{ex}}}{M_{\text{S}} \mu_{0}}) k^2 + H_{0} & \frac{i 2\pi f}{\gamma \mu_{0}} & -M_{\text{S}} & 0& 0\\
i 2 \pi f \mu_{0}  & 0 & i 2 \pi f \mu_{0} & 0& \partial_{y}+i k\\
0& i 2 \pi f \mu_{0} & 0 & i 2 \pi f \mu_{0} & -\partial_{x} \\
0  & 0 & -\partial_{y} - i k &  \partial_{x}& -\sigma
\end{array}}
\right) \left( 
\scalemath{0.75}{
\begin{array}{c}
m_{x}^\prime  \\
m_{y}^\prime \\
h_{x}^\prime  \\
h_{y}^\prime  \\
e_{z}^\prime
\end{array}}\right)=0. \label{Eq:FEM}
\end{equation}
\end{widetext}
This set of equations is solved with the use of finite element method implemented in COMSOL MultiPhysics software. From the solutions we obtain dispersion relation of SWs and vectors which represent the spatial distribution of dynamical components of  $m^\prime_{x}$,
$m^\prime_{y}$, $h^\prime_{x}$, $h^\prime_{y}$  and $e^\prime_{z}$. Since the structure is periodic in the real space with periodicity of $a$, the solutions (frequencies of SW) possess the periodicity in the 1D reciprocal space characterized by the period $2 \pi$/$a$ with artificial folding to the first Brillouin zone. In order to investigate SWs in a uniform film (so called empty lattice model) we calculate the SW frequencies for the wavevectors from the first Brillouin zone and eliminate the artificial solutions from higher bands. The lattice constant $a=200$ nm was chosen in all calculations, so the computational unit cell is not large and the higher branches of empty lattice model do not cross with the first branch in the considered frequencies in this paper.
%On the interfaces between MC and dielectric the solutions satisfy electromagnetic boundary conditions and exchange boundary conditions: $\frac{\partial m^\prime_{i}}{\partial y}=0$ ($i  = x, y$). On the interfaces between magnetic materials $m^\prime_{i}$ and $\frac{A}{M_{\text{S}}} \frac{\partial m^\prime_{i}}{\partial x} $ are continuous functions ($M_{\text{S}}$ is a saturation magnetization). The interactions with electromagnetic, acoustic waves or microwave currents can be added by generalizing the system of equation presented above.

\subsection{Semi-analytical method}\label{Sec:SA}
In semi-analytical model the dynamic magnetic field induction:\cite{Berg}
\begin{equation}
\textbf{ b}=\mu_{0}(\textbf{h}+\textbf{m})
\label{eq 5}
\end{equation}
 is represented in terms of the vector potential function $\textbf{A}$:
\begin{equation}
\textbf{b} = \nabla \times  \textbf{A}= (\partial_{y} A_{z},-\partial_{x} A_{z},\partial_{y} A_{x}-\partial_{x} A_{y}).
\label{eq 6}
\end{equation}
With this vector potential the Gauss low $\nabla \cdot \textbf{b} =0$ is identically fulfilled. From LL equation we can define magnetic permeability tensor $\hat{\mathbf{\mu}}$ and express $\mathbf{h}$ field in terms of the $\mathbf{b}$ field. When $A_{\text{ex}}=0$ it reads:
\begin{equation}
\textbf{ h} = \hat{\mathbf{\mu}}^{-1} \textbf{ b} = \left(\begin{array}{ccc}
\frac{1+\kappa}{(1+\kappa)^2-\nu^2} & \frac{i \nu}{(1+\kappa)^2-\nu^2}  & 0\\
 -\frac{i \nu}{(1+\kappa)^2-\nu^2}  & \frac{1+\kappa}{(1+\kappa)^2-\nu^2} & 0\\
0&0&\frac{1}{2}
\end{array}\right) \textbf{ b},
\label{eq 7}
\end{equation}
where:
\begin{equation}
\kappa= \frac{\Omega_{H}}{\Omega_{H}^2-\Omega^2}, \;
\nu=\frac{\Omega}{\Omega_{H}^2-\Omega^2}
\end{equation}
and 
\begin{equation}
\Omega= \frac{2 \pi f}{\gamma \mu_{0} M_{\text{S}}}, \;
\Omega_{H}= \frac{H_{0}}{M_{\text{S}}}.
\end{equation}
 From Eqs.~(\ref{eq 5})-(\ref{eq 7}) one can see that $h_{x}$, $h_{y}$, $m_{x}$ and $m_{y}$ are defined by $A_{z}$ component of the vector potential. From  equations (\ref{eq:2}) and (\ref{eq:3}) we obtain following equation for $A_{z}$:
\begin{equation}
\left( \frac{1+\kappa}{(1+\kappa)^2-\nu^2} \right) \nabla^2 A_{z}= i 2 \pi f \sigma \mu_{0} A_{z}.\label{Eq:Az}
\end{equation}

The total solution of the Eq.~(\ref{Eq:Az}) can be found from the solutions in regions I to V imposed into boundary conditions at interfaces along $x$ axis and at infinity.  The solution of $A_{z}$ in the region IV (metal) can be written in the following form:
\begin{equation}
A_{z}(x,y)=(a_{4}\text{e}^{k_{\text{M}}x}+b_{4}\text{e}^{-k_{\text{M}}x})\text{e}^{-i k y}, \label{Eq:Az_M}
\end{equation}
where $a_{4}$ and $b_{4}$ are constants. $k$ is a wavenumber of wave propagating along $y$ axis and can take any real value. The wavenumber along $x$ axis in the metal, $k_{\text{M}}$ is derived from Eq.~(\ref{Eq:Az}) with the use of Eq.~(\ref{Eq:Az_M}):
\begin{equation}
k_{\text{M}}=\sqrt{k^2+\frac{2 i}{\delta_{0}^2}},\label{Eq:kM}
\end{equation} 
where
\begin{equation}
\delta_{0}=\sqrt{\frac{1}{ \pi f \mu_{0} \sigma}} \label{Eq:delta}
\end{equation}
is a frequency dependent skin depth of the metal.
For the remaining regions (regions I - III and V, where $\sigma \equiv 0$) the $A_{z}$ can be taken in the form:
\begin{equation}
A_{z}(x,y)=(a_{n}\text{e}^{k x}+b_{n}\text{e}^{-k x})\text{e}^{-i k y}, \label{Eq:Az_I}
\end{equation}
where $a_{n}$ and $b_{n}$ are constants. $n$ denotes the region: I, II, III or V. In region I and V we expect that the any $\psi \prime$ function decay for  $x \rightarrow -\infty$ and $x \rightarrow \infty$, thus $b_{1}=a_{5}=0$.

 By imposing the solutions (\ref{Eq:Az_M}) and (\ref{Eq:Az_I}) into electromagnetic boundary conditions, it is a continuity of tangential component of the magnetic field $h_{y}$ and the normal component of the magnetic induction field $b_{x}$ the secular equation is obtained. This equation is solved using Newton iterative method. From the condition of existence of nontrivial solutions the dispersion relation of SW in magnetostatic approximation can be obtained as well.

\section{Results}

\begin{figure}[!ht]
\includegraphics[width=0.4\textwidth]{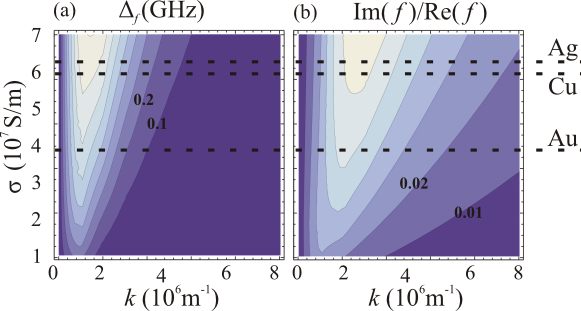}
\caption{(Color online) Color map of the metallization effect $\Delta_{f}$ and the inverse of FOM $\left[ 1/\text{FOM} \right]$ as a function of the wavenumber $k$ and the conductivity of the metallic overlayer $\sigma$ is shown in (a) and (b), respectively. Horizontal black dashed lines mark the conductivity of a few common metals: Ag, Cu and Au.}
\label{sigma}
\end{figure}

In Fig.~\ref{sigma}(a) we present the effect of a metallization ($\Delta_f$) of the SWs propagating in thin film  as a function of the conductivity of the metal overlayer and the wavenumber of a SW calculated with FEM. We define the measure of the metallization effect as a difference between frequency of SW propagating in positive $y$ direction in the ferromagnetic film with metallic overlayer and  with dielectric surroundings from both sides, i.e., $\Delta_{f} (\sigma, k)= f(\sigma, k) - f(0,k)$.\footnote{This definition of $\Delta_f$ is related also to the difference between frequencies of SW propagating in positive and negative direction in the same structure [i.e. to the nonreciprocity strength introduced at the end of the Sec.~\ref{Sec:SB}]. It is because the propagation of the wave with amplitude localized at the surface with dielectric is weakly affected by the presence of the metal on the opposite side.\cite{mru}} The calculations were performed for the ferromagnetic film with the following parameters: $M_{\text{S}}=1.2 \times 10^6$ A/m, $d=30$ nm, $t = 0$, $D = 150$ $\mu$m, $A_{\text{ex}}=2.8 \times 10^{-11}$ J/m and with the in-plane bias magnetic field $\mu_0 H_{0}=0.1$ T.\footnote{We have verified by direct comparison of the FEM results with the semi-analytical results performed with neglecting exchange interactions that the value of an exchange constant does not influence  results in the considered here wavenumbers and frequencies.}  Conductivity values of some common metals (Ag, Cu and Au) at room temperature are indicated by horizontal dashed lines in Fig.~\ref{sigma}. We see that for Ag, a metal with highest conductivity, $\Delta_f$ has its maximum at $k\approx 1 \times 10^6$ 1/m and $\Delta_f$ extends the value of 0.8 GHz. For other metals $\Delta_f$ is smaller but always its maximal value is for the similar value of $k \approx 10^6$ m$^{-1}$.

In Fig.~\ref{sigma}(b) the reversal of the figure of merit (FOM) defined as $\frac{1}{\text{FOM}}=\frac{\text{Im}(f)}{\text{Re}(f)}$ in dependence on $\sigma$ and $k$ is shown. This function describes the attenuation of SW, it is an attenuation induced by metallization of the overlayer, because the intrinsic damping is not considered here. Thus for parameters where the dielectrics are on both sides of the ferromagnetic film FOM $\rightarrow \infty$  ($1/\text{FOM} \rightarrow 0$). The FOM is an important parameter regarding the potential applications, because in order to propagate signal for large distances, SWs need to have a long life time,  i.e., the real part of frequency shell to be much higher than the imaginary part: Re$(f)>$ Im$(f)$.\cite{Saratov2} It means that  $1/\text{FOM}$ has to be smaller than 1. In Figs.~\ref{sigma}(a) and (b) we can see that $\Delta_f$ and 1/FOM have similar dependence on $\sigma$ and $k$ but the maximal value of 1/FOM is shifted to higher wavenumbers as compared to $\Delta_f$. 

To explain presented above results of FEM calculations we will perform now  an analysis of the dispersion relation and the fields emitted by the SW excitations outside of the magnetic film with the use of semi-analytical method described in Sec.~\ref{Sec:SA}. The structure with conductivity $\sigma= 6 \times 10^7$ S/m (i.e., close to the value of Cu or Ag) and parameters described above will be considered as a base for further analysis of this paper. Any variation of these parameters will be indicated. 

\subsection{Approximate analytical solution}\label{Sec:approximate}

\begin{figure}[!ht]
\includegraphics[width=0.45\textwidth]{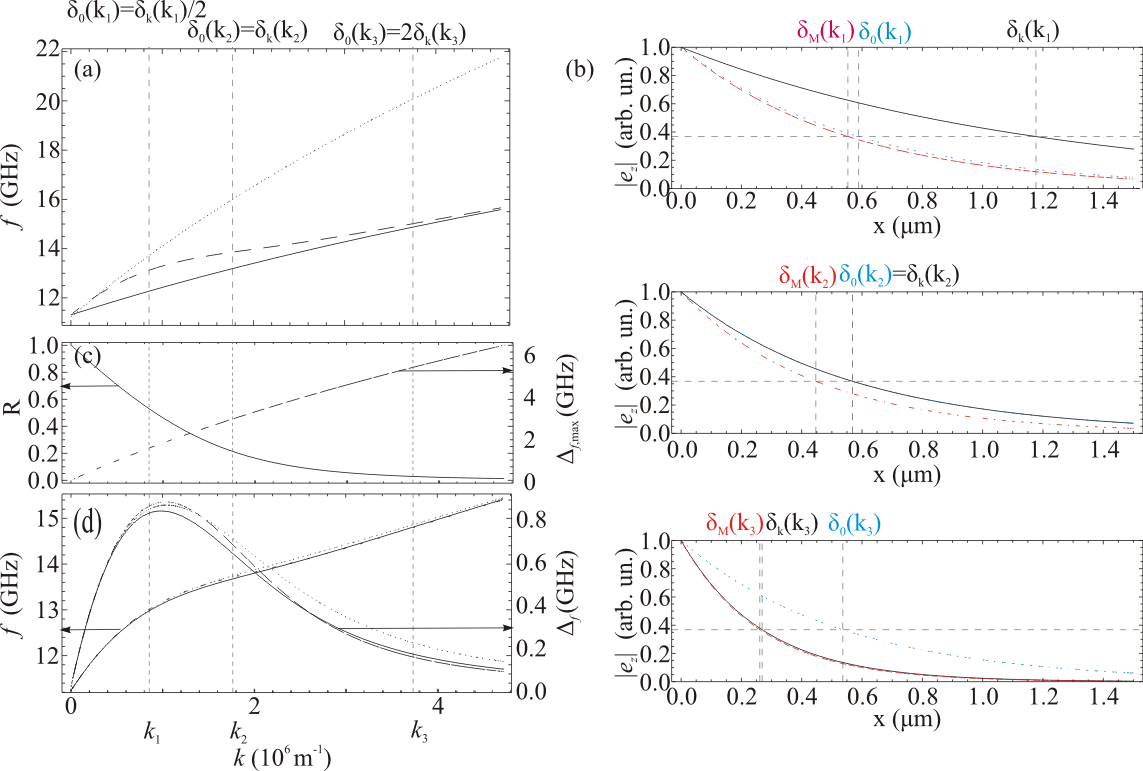}
\caption{(Color online)  (a) Dispersion relation of DE wave in ferromagnetic film of $d=30$ nm thickness and saturated by the external magnetic field $\mu_0 H_{0}=0.1$ T. Solid line shows calculation's result for the film with both dielectric  surroundings, dotted line for a film covered from one side with PEC and dashed line with metal with $\sigma= 6 \times 10^7$ S/m conductivity. (b) Amplitude of the $z$ component of the electric field decaying with increasing distance from the surface of the ferromagnetic film as $\exp{(-x/\delta_{k})}$---solid line, $\exp{({-x/ \delta_{\text{M}}})}$---dashed line and as $\exp{(-x/ \delta_{0})}$---dotted line are shown for three values of the wavenumber. It is for $k_{1}=0.85 \times 10^6$ m$^{-1}$, $k_{2}=1.76 \times 10^6$ m$^{-1}$ and $k_{3}=3.72 \times 10^6$ m$^{-1}$ marked also by vertical lines in (a), (c) and (d). (c)  $R$ and $\Delta_{f,\text{max}}$ as a function of the wavenumber are shown with solid and dashed line, respectively. (d) The approximate and exact values of $f$ and $\Delta_f$ in dependence on the $k$. Solid line shows $f_{\text{app}}(k)$ calculated from Eq.~(\ref{Eq:f_aver}) with $\delta_0$ estimated  with Eq.~(\ref{Eq:delta_DE}),  dot-dashed line with $\delta_0$ estimated  with Eq.~(\ref{Eq:delta_av}), and dashed line shows the $f(k)$ obtained in semi-analytical model. }
\label{Fig2}
\end{figure}

Results of calculations for the structure with parameters specified above, where metal is in direct contact with ferromagnetic film and its thickness is much larger than the skin depth of metal are presented in Fig.~\ref{Fig2}(a). The three dispersion relations show: the results of semi-analytical calculations (which coincide with the FEM results)--dashed line; the dispersion of SWs for the same film but in contact with dielectric--solid line and in contact with perfect electric conductor (PEC)--dotted line. We can see, that at small wavenumbers ($k < k_1$) the dispersion of DE wave with metallic overlayer of finite conductivity (dashed line) follows the dispersion in ferromagnetic film in contact with PEC (dotted line), while for larger wavenumbers ($k$ larger than $k_3$) it follows the dispersion relation in the film with both dielectric surroundings (solid line). 

In order to understand the range of wavenumbers for which the SW dispersion is under influence of a metal overlayer with finite conductivity we will analyze penetration depth of electromagnetic field into a metal and compare it with the penetration depth in a dielectric. Decay of the electric field associated with the SW excitation is exponentially depended on its wavevector and characterized by the penetration depth. In the case of ferromagnetic film surrounded by a dielectric the penetration depth is $\delta_{k}=1/k$. If the magnetic film is surrounded by a metal, an additional attenuation of the field is present and this is described by metal's skin depth $\delta_{0}$ [Eq.~(\ref{Eq:delta})]. The total penetration depth in this case is $\delta_{\text{M}}$ and is related to the metal wavenumber introduced in Eq.~(\ref{Eq:kM}): $\delta_{\text{M}}=1/ k_{\text{M}}$. Which of the two decay contributions, $k= 1/\delta_{k}$ or $1/ \delta_{0}$ to $k_{\text{M}}$ is more important, will determine the influence of a metal on the SW excitation.

In the subsequent figures in Fig.~\ref{Fig2}(b) we plot $\left|e_z (x)\right|$ (dashed line)  along $x$ axis in the space occupied by the metal ($x > 0$) for three values of the wavevector. In the top figure for $k_{1}=0.85 \times 10^6$ m$^{-1}$, in the middle figure for $k_{2}=1.76 \times 10^6$ m$^{-1}$ and in the figure at the bottom for $k_{3}=3.72 \times 10^6$ m$^{-1}$. These wavenumbers are marked by vertical lines also in Fig.~\ref{Fig2}(a). In each plot $e_z(x) \propto \exp{({-x/ \delta_{\text{M}}})}$ dependence (dashed line) is decomposed onto two components: a decay due to nature of DE's SW ($\exp{(-x/\delta_{k})}$, solid line) and due to finite conductivity attenuation in metal ($\exp{(-x/ \delta_{0})}$, dotted line). All functions are calculated with the use of semi-analytical method and normalized so that $\left|e_{z} \right|$=1 at the surface of the ferromagnetic film (at $x=0$). The analytical dependencies are in perfect agreement with results obtained from FEM calculations, not shown here. The 1/e value of $e_z$ is marked by the horizontal dashed line. We see that in the first case decay is determined by a decay in metal because $\delta_{\text{M}}(k_1) \approx \delta_{0}(k_1) < \delta_{k}(k_1)$, in the second case the skin depth of metal is close to the penetration depth of DE wave ($\delta_{0}(k_2) \approx \delta_{k}(k_2)$), this is intermediate case where both sources of attenuation have the same contribution to $\delta_{\text{M}}$.  Whereas in the third case $\delta_{\text{M}}(k_1) \approx \delta_{k}(k_3) < \delta_{0}(k_3)$ and the total attenuation is due to localization of the DE excitation at this wavelength, i.e., the penetration depth into dielectric is shorter than the skin depth of metal, an effect of metal is not visible. This analysis allows to define three wavenumber regions: i) $k<k_{1}$, the frequency is strongly influenced by metal, it corresponds to a dispersion of ferromagnetic film in contact with PEC. ii) $k_{1}<k<k_{2}$, the influence of metal is visible, but not as strong as magnetic film would be in contact with PEC, but here the maximal influence of metallization (also a maximal nonreciprocity) on SW is expected, and iii) $k_{3}<k$, the effect of metal is negligible.

We can estimate the quantitative contribution of a metal attenuation to the total attenuation of SW in the region IV [Fig.~\ref{geometry}] by calculating the ratio, $R$, of the electric field attenuation in metal due to the conductivity (i.e., the electric field attenuation due to the SW excitation divided by the total electric field attenuation): 
\begin{equation}
R=1- \frac{ \int^{\infty}_{0}{\left|\text{e}^{- k x}\right| dx}}{ \int^{\infty}_{0}{\left|\text{e}^{-k_{\text{M}} x}\right| dx}}= \frac{\delta_{\text{M}}-\delta_{k}}{\delta_{\text{M}}}.
\end{equation} 
The function $R$ changes from 0 to 1; $R$ close to 0 means that almost all the electric field attenuation  is due to decay of the SW excitation, close to 1 means that almost all electric field attenuation is due to the presence of metal. $R$ as a function of SW wavenumber is plotted in Fig.~\ref{Fig2}(c) with solid line. Function is monotonic and changes from 1 at $k$ = 0 to 0 at $k \rightarrow \infty$, as expected. 
In order to obtain the approximate value of the metallization effect, we will use the  frequencies in the structure with PEC and without metal, thus it will define $\Delta_{f,{\text{max}}}$, i.e.,  a maximal available metallization effect. We will use here analytical formulas for SW dispersion of an uniform film surrounded by dielectrics and in direct contact with PEC, $f_{\text{DE}}(k)$ and $f_{\text{PEC}}(k)$, which are defined in Refs.~[\onlinecite{DE}] and~[\onlinecite{Sachardi}], respectively. Thus, $\Delta_{f,{\text{max}}}(k) \equiv f_{\text{PEC}}(k)-f_{\text{DE}}(k)$ and its dependence on $k$ is shown in Fig.~\ref{Fig2}(c) with dashed line. The approximate strength of the metallization effect, $\Delta_{f,\text{app}}$ is then given by:
\begin{eqnarray}
\Delta_{f} &\approx& \Delta_{f,\text{app}} = \Delta_{f,{\text{max}}} R= (f_{\text{PEC}}-f_{\text{DE}}) R \nonumber \\
& =&(f_{\text{PEC}}-f_{\text{DE}}) \frac{\delta_{\text{M}}-\delta_{k}}{\delta_{\text{M}}}. \label{Eq:Delta_aver}
\end{eqnarray}
The approximate real part of SW's frequency can be also calculated with approximate formulas: 
\begin{eqnarray}
f &=& f_{\text{DE}}+\Delta_{f} \approx f_{\text{app}} \nonumber \\
&= & f_{\text{DE}}+(f_{\text{M}}-f_{\text{DE}}) \frac{\delta_{\text{M}}-\delta_{k}}{\delta_{\text{M}}}, \label{Eq:f_aver}
\end{eqnarray}
where $\delta_{M}$ depends on $\delta_{0}$ [Eq.~(\ref{Eq:kM})] and so it is still a function of $f$. However this frequency can be approximated, we propose to use two approaches to estimate $f$ in Eq.~(\ref{Eq:kM}):\\
 i)   from  the analytical solution for DE mode in ferromagnetic film surrounded by dielectrics: 
\begin{equation}
\delta_0 \approx \delta_{\text{DE}}=\sqrt{\frac{2}{  2 \pi \mu_{0} f_{\text{DE}} \sigma}}, \label{Eq:delta_DE}
\end{equation}
ii) from the averaged frequency of the film with PEC and with dielectrics:
\begin{equation}
\delta_0 \approx \delta_{\text{avrg}}=\sqrt{\frac{2}{  2 \pi \mu_{0} \frac{f_{\text{DE}}+f_{\text{PEC}}}{2} \sigma}}. \label{Eq:delta_av}
\end{equation}
%and this will be used to calculate the $f_{\text{app,DE}}$ from Eq.~(\ref{Eq:f_aver});This will be used to calculate $f_{\text{app,avrg}}$ also from Eq.~(\ref{Eq:f_aver}).

In Fig. \ref{Fig2}(d) we plot both, $\Delta_{f,\text{app}}(k)$ and $f_{\text{app}}(k)$ for the structure in contact with metal, obtained from Eqs.~(\ref{Eq:Delta_aver}) and (\ref{Eq:f_aver}) with using skin depths Eqs.~(\ref{Eq:delta_DE}) and (\ref{Eq:delta_av}) (solid and dotted lines, respectively) and compare them with numerical solutions ($\Delta_{f}(k)$ and $f(k)$ obtained with semi-analytical method, marked by dashed line). We could expect that the choice of $f$ function in $\delta_{0}$ may influence results. In fact, we observe the differences but these are very small and almost invisible in frequency dependence on the wavevector due to frequency scale used in figure [the left scale in Fig.~\ref{Fig2}(d)]. However, we can see the discrepancy between numerical results and approximate solutions in the plot of the metallization (the right scale) for $k > k_1$. At small $k$  numerical and both approximate solutions are very close to each other, this shows that the function $\delta_0$ is approximated correctly at these wavenumbers. The differences  between numerical solution of $\Delta_f(k)$ and $\Delta_{f,\text{app}}(k)$ based on DE approximation Eq.~(\ref{Eq:delta_DE}) and the average value Eq.~(\ref{Eq:delta_av}) (both approximate values coincide) increases up to wavenumber where $\Delta_f$ reach maximal value.  It is since the frequency used in $\delta_{0}$ is becoming significantly different from right solution. This discrepancy is eliminated when iterations of frequency are used. For large $k$ the $\Delta_{f,\text{app}}$ based on DE approximation is again close to numerical solution, while $\Delta_{f,\text{app}}$ based on average value is higher from numerical solution and this difference saturate with increasing $k$. Also the influence of error in $\delta_0$ due to frequency approximation is smaller at small $k$ than at large $k$, since the value of $R$ is larger at small $k$.

\subsection{Maximizing the influence of metal}\label{Sec:SB}

\begin{figure}[!ht]
\includegraphics[width=0.4\textwidth]{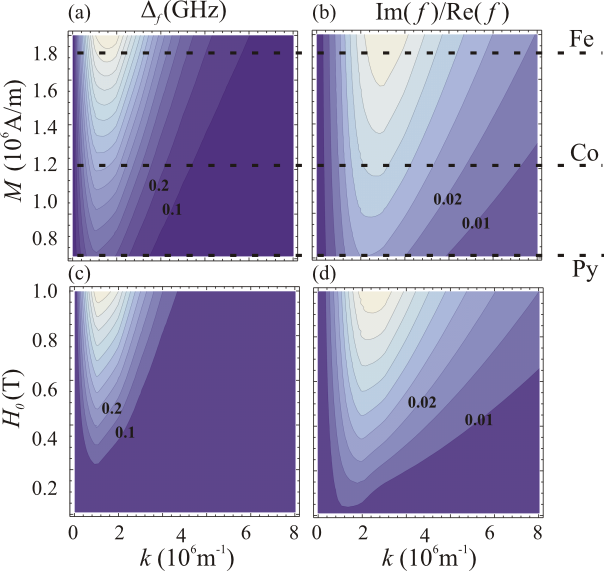}
\caption{(Color online) The strength of a metallization effect $\Delta_f$  as a function of (a) $k$ and $M_{\text{S}}$, (c) $k$ and $H_{0}$. %and (e) $k$ and $d$.
 The inverse of FOM [$1/$FOM] as a function of (b) $k$ and $M_{\text{S}}$, (d) $k$ and  $H_{0}$.  In (a) and (b) the dashed horizontal lines mark saturation magnetization of Py, Co and Fe.}
\label{3}
\end{figure}

Based on the discussion from the previous subsection we can point at parameters which will be important in order to increase the effect of metallization (to decrease skin depth of metal $\delta_{0}$) on the dispersion of SWs and so, to increase the strength of nonreciprocity. Since now we will use solely FEM method for analysis. According to Eq.~(\ref{Eq:delta}) the skin depth is proportional to the square root from the inverse of frequency:
$$\delta_{0} \propto \sqrt{\frac{1}{f}},$$
thus increasing the frequency at fixed $k$ will lead to shorter skin depth in metal and the effect of metallization will be stronger (higher attenuation due to electron screening and higher value of function $R$). One can expect that the frequency at fixed $k$ will increase by:
\begin{itemize}
\item{} applaying higher bias magnetic field,
\item{} using material with higher magnetization,
\item{} increasing the thickness of the magnetic film.\footnote{In this paper we limit our investigation to the thin ferromagnetic films ($d \le$ 30 nm) when the effect of its finite conductivity on the SW propagation is negligibly as was proved by a number of experiments on thin Py films. In the thicker ferromagnetic films the eddy currents generated in ferromagnetic film can influence the SW dispersion.\cite{Mills}}
\end{itemize}
The results of numerical calculations confirm the predictions. In Figs.~(\ref{3}) (a) and (c) the $\Delta_f$ is plotted in two-dimensional maps as a function of $(M_{\text{S}},k)$ and $(H_0,k)$, respectively. 

\begin{figure}[!ht]
\includegraphics[width=0.4\textwidth]{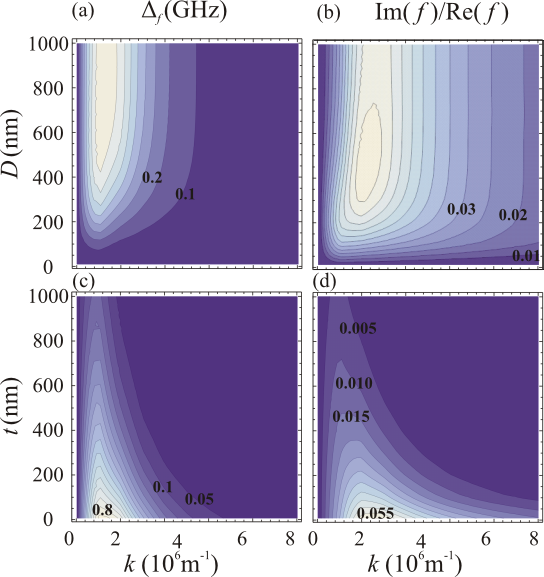}
\caption{(Color online) $\Delta_f$ as a function of (a) $k$ and the thickness of the metallic layer $D$, (c) $k$ and the separation between metal and ferromagnetic film $t$.  1$/$FOM as a function of (b) $k$ and $D$, and (d) $k$ and $t$.}
\label{4}
\end{figure}
In Figs.~(\ref{3}) (b) and (d) the reverse of the FOM function is plotted for the same dependence and it can be directly compared with Gilbert damping factor, since $1/\text{FOM} \approx \alpha$.\cite{Javier}  For the structure with parameters taken from Fig.~(\ref{Fig2}) (a) the imaginary part of the frequency is always much smaller than the real part and $1 / \text{FOM}$ reaches the value 0.06 ($\text{FOM} \approx 16.7$), which is of the same order as a Gilbert damping in many ferromagnetic materials.\cite{damping, damping 2} In other considered here parameters we observe increase of $1/\text{FOM}$ with increase of $M_{\text{S}}$ and $d$, but a decrease with an increase of $H_{0}$. 

In Fig.~(\ref{3}) (a) the magnetization saturation of Fe, Co and Py are indicated. The effect of metallization is negligible for thin films made of materials with low magnetisation, it is already for Py. Thus, the new materials with high magnetization, like a CoFeB alloys proposed recently for magnonics applications\cite{Yu12,CoFeB} or Heusler alloys \cite{Tru10,Seb13,Dub13} might be useful for a fabrication the structures with high nonreciprocal effect and high velocity of SW with reasonable value of damping.

In Figs.~(\ref{4}) (a)-(d)  the  strength of the metallization effect and 1/FOM in dependance on the thickness of the metal film $D$, a metal separation from the magnetic film $t$ and a wavenumber are presented. For $D$ dependance we can observe a saturation of $\Delta_f$ at values of order of skin depth in metal ($\delta_{0}$ is usually in the range of 550-620 nm). However,  a small variation of 1/FOM is still visible for thicker metallic film. The interesting point is that the region on the $(D,k)$ plane, where 1/FOM $> 0.063$ has widest range of $k$ when $D \approx \delta_{0}$. The $t$ dependence of $\Delta_f$ and 1/FOM is opposite to dependence on $D$ described above. Both functions decrease with increasing $t$, however they have a maximum in dependence on $k$, this maximum exists for SWs with small $k$ wavevectors, i.e., when $\frac{1}{k} > t+ \delta_{0}$ and shifts slightly to the smaller $k$ with increasing $t$.  

In addition to the magnetization, external field, metallic film thickness and metal separation, the change of the thickness of a ferromagnetic film  will also have impact on nonreciprocity effect at given wavevector. With the increase of thickness, the group velocity of DE wave increases. At given wavevector the frequency of SW is higher for thicker sample, thus the penetration depth of the excitation decreases and an influence of metal is stronger. However, with the increase of ferromagnetic film thickness, the influence of a conductivity of the ferromagnetic film on the dispersion of SWs becomes not negligible. Moreover, the damping also increases limiting potential applications of such structures. Thus we have limited our investigations to a thin ferromagnetic films. The thickness dependance on nonreciprocity strength might be also considered in nonconductive magnetic films, e.g., in YIG, then the thicker sample is, the stronger effect is expected.\cite{De_Wames, De Wames 4}

\begin{figure}[!ht]
\includegraphics[width=0.5\textwidth]{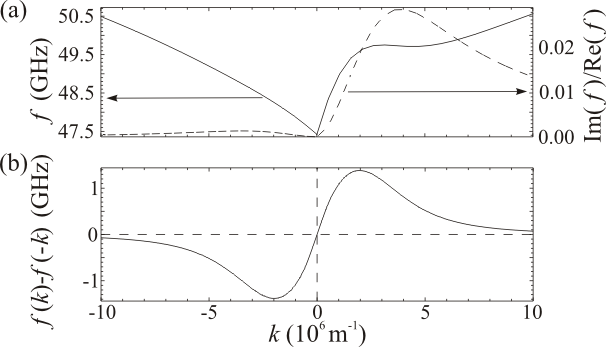}
\caption{(Color online) The dispersion relation (solid line) and the 1$/$FOM (dashed line) of DE's SW in the 30 nm thick CoFeB film with a Cu overlayer of 500 nm thickness. The following parameters were assumed in the numerical calculations: $\mu_0 H_{0}=1$ T, $M_{\text{S}}=1.43  \times 10^6$ A/m, $A = 2.8 \times 10^{-11}$ J/m and $\sigma = 6 \times 10^7$ S/m. (b) The nonreciprocity strength for CoFeB thin film in direct contact with the Cu overlayer.}
\label{5}
\end{figure}

There is also another factor, not considered in this paper, which can contribute to the nonreciprocal properties of SW in thin films, this is an exchange interaction. The exchange interactions result in appearing of $~ A k^2$ term in the magnetostatic SW dispersion relation.\cite{Stancil_b}  Thus high exchange constant should lift the frequency up at the fixed $k$ and decrease $\delta_{0}$, so the metallization can be visible at larger $k$ values. However, as we have verified numerically, to observe this effect the value of $A$ should be at least two orders of magnitude higher than that of ferromagnetic materials considered in this paper.

On the basis of  analysis presented above we can propose the material and the structure suitable for an observation of the nonreciprocal dispersion relation of SWs in thin ferromagnetic film with currently available resolution of the BLS setups.\cite{Gubbiotti10} The high magnetization material was chosen, it is  CoFeB alloy,\cite{CoFeB} to be studied at relatively high external field,  $\mu_0 H_{0}=1$ T, and covered with the thick film of Cu. In Fig.~(\ref{5})(a) we show the dispersion relation [solid line] and $1/ \text{FOM}$ [dashed line] for a 30 nm thick film with high nonreciprocal dispersion of SWs at wide wavevector range. The nonrecirpocity strength defined as the difference between frequency of SW propagation in $+k$ direction and $-k$ is shown in Fig.~(\ref{5})(b).  The maximal nonreciprocity is found at $k \approx 2.0 \times 10^6$ 1/m with $f(k) - f(-k) \approx 1.3$ GHz. The nonreciprocity higher than 0.2 GHz can be observed for wavevectors in the range of $ 7.0 \times 10^6$ 1/m. There is also interesting effect related to the SW attenuation, it is the big noreciprocity in an attenuation of SW due to the presence of metal. SWs propagated in $-k$ direction are almost undamped, while in $+k$ are attenuated with maximum 1/FOM at $4 \times 10^6$ 1/m. However this maximum does not coincide  with the maximal value of a nonreciprocity strength shown in Fig.~(\ref{5})(b).   

\section{Conclusions}\label{Sec:Concl}
We have presented the results of our investigation of the influence of the metallic overlayer on the SW dispersion relation in ferromagnetic thin films. By analyzing the electric field component within the finite-conductivity metallic layer we have estimated the extent of the nonreciprocity in such structures versus various parameters, including the magnetic saturation of the ferromagnetic film, the conductivity of the metallic layer, its thickness and distance from the ferromagnetic film, and the in-plane external magnetic field.  Using an approximate analytical formula for the SW frequency, we have defined a structure in which the effect of the metallization is significant in a wide wave-vector range and should be measurable by standard experimental techniques suitable for the measurement of SWs.

We have shown that the results obtained by the FEM in the present study agree with the semi-analytical data. A major advantage of the FEM is its applicability, in the present formulation, to various structures, including magnonic crystals,\cite{mru} multilayered structures, ferromagnetic films with layers of various conductivity, films with corrugated surfaces or in contact with a metallic grid. These additional structural variations might further increase the nonreciprocal effect.

The results presented here can be of use for the development of devices exploiting the nonreciprocity of the SW dispersion relation. The proposed structure, based on a CoFeB thin film, can be further explored for magnonic applications\cite{Kruglak10} by periodic patterning of the ferromagnetic film or patterning of the metallic cover layer. The nonreciprocal properties can provide additional tunability of the magnonic band structure in variety of magnonic crystals\cite{Klos12,Mamica12,Neusser11} for various applications. Its magnonic band structure can be of use e.g. for the design of high-sensitivity magnetic field sensors,\cite{Inoue} miniaturized microwave isolators or circulators, or other elements vital to modern microwave technology. Moreover, the fabrication of the structure proposed in this paper is relatively easy, and its nonreciprocity can be combined with the reprogrammability of the magnonic band structure of an array of ferromagnetic stripes created from the structure proposed.\cite{Topp10,Huber13}

\begin{acknowledgments}
The research leading to these results has received funding from
the European Community's Seventh Framework Programme
Grant Agreements no 247556 (People)--NoWaPhen and from National Science Centre of Poland project no DEC-2012/07/E/ST3/00538. 
\end{acknowledgments}

\end{document}